# Astro2020 Science White Paper

# State of the Profession: Intensity Interferometry

**Thematic Areas:** ☒ Ground Based Project   ☐ Space Based Project

☒ Infrastructure Activity   ☒ Technological Development Activity

☒ State of the Profession Consideration

**Principal Author:**
Name: D. Kieda
Institution: University of Utah
Email: dave.kieda@utah.edu
Phone: 801-581-6926

**Abstract**:

Recent advances in telescope design, photodetector efficiency, and high-speed electronic data recording and synchronization have created the observational capability to achieve unprecedented angular resolution for several thousand bright (m< 6) and hot (O/B/A) stars by means of a modern implementation of Stellar Intensity Interferometry (SII). This technology, when deployed on future arrays of large diameter optical telescopes, has the ability to image astrophysical objects with an angular resolutions better than 40 μ arc-sec. This paper describes validation tests of the SII technique in the laboratory using various optical sensors and correlators, and SII measurements on nearby stars that have recently been completed as a technology demonstrator. The paper describes ongoing and future developments that will advance the impact and instrumental resolution of SII during the upcoming decade.

**Co-authors:** (names and institutions)
Gisela Anton, ECAP, University of Erlangen
Anastasia Barbano, Universite de Geneve
Wystan Benbow, Center for Astrophysics | Harvard & Smithsonian
Colin Carlile, Lund University
Michael Daniel, Center for Astrophysics | Harvard & Smithsonian
Dainis Dravins, Lund University
Sean Griffin, University of Maryland
Tarek Hassan, DESY-Zeuthen
Jamie Holder, University of Delaware
Stephan LeBohec, University of Utah

Nolan Matthews, University of Utah
Theresa Montaruli, DNPC, Universite de Geneve
Nicolas Produit, Universite de Geneve
Josh Reynolds, Cork Institute of Technology
Roland Walter, *DNPC,* Universite de Geneve
Luca Zampieri, INAF-Osservatorio Astronomico di Padova

**Dedication**

*The authors of this APC white paper dedicate this contribution in memory of our colleague, Dr. Paul Nuñez, who passed away in an unfortunate mountain climbing accident in June 2019. Dr. Nuñez is widely recognized for his foundational work in the development of astronomical SII imaging.*

# The Impact of Multi-kilometer Baseline Visible Band Interferometry

Imaging is one of the basic tools of astronomy. Our ability to understand an astronomical object is often driven by our ability to resolve it and study its dimensions and morphology. Bright stars typically have angular diameters of the order of one milli-arcsecond (mas), and revealing details requires angular resolutions measured in micro-arcseconds (μas). In the visible light waveband, an angular resolution below 100 μas can only be realized by interferometers with telescopes separated by hundreds to thousands of meters.

Stellar Intensity interferometry (SII) exploits a second-order effect of light waves, viz. how instantaneous fluctuations in intensity correlate between spatially separated telescopes (Figure 1). The technique was developed by Robert Hanbury Brown and Richard Q. Twiss, who built the Narrabri Stellar Intensity Interferometer (NSII) using twin 6.5 m diameter telescopes movable along a 188 m diameter circular railroad track at Narrabri, New South Wales, Australia [1]. SII possesses the great observational advantage of being practically insensitive to both atmospheric turbulence and to telescope optical imperfections, allowing very long baselines, as well as observations at short optical wavelengths. Telescopes are connected only electronically, with no need to directly combine the photons they detect. Noise is therefore introduced by the electronics, with timescales of nanoseconds, and by light coherence lengths (centimeters or even meters), rather than by wave period and wavelengths (with scales of femtoseconds and nanometers, respectively). A time resolution of 3 ns, say, corresponds to about one-meter light-travel distance, and thus atmospheric path-lengths and telescope imperfections only need to be controlled within some reasonable fraction of one meter.

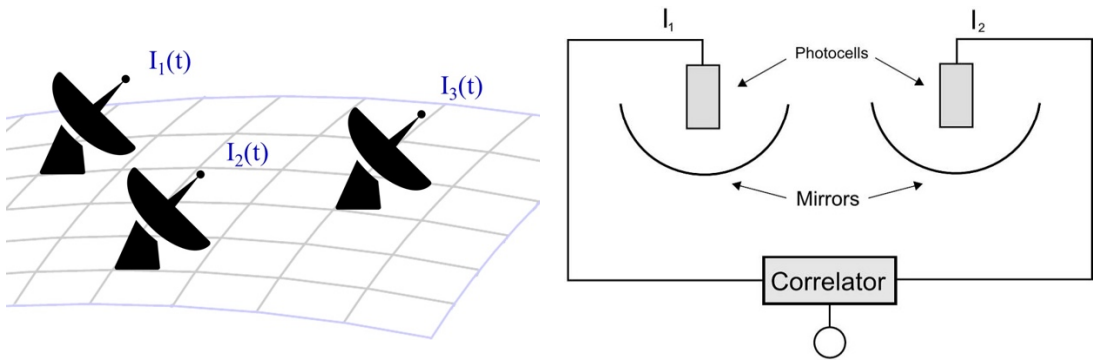

Figure 1: **Left**: Spatially separated telescopes observe the same source, and the measured time-varying intensities $I_1(t)...I_3(t)$ are electronically cross-correlated between different pairs of telescopes. **Right**: Basic components of a two-telescope intensity interferometer.

The fraction of a meter pathlength requirement for SII removes the two main obstacles for extending optical interferometers over longer baselines and combining a greater number of telescopes. Atmospheric turbulence is of no practical concern and, since the signals are electronic only, these may be freely combined, copied, stored or processed among any number of telescopes, on- or off-line, enabling visible waveband observations using a detection method similar to long-baseline radio interferometric arrays.

However, the SII method is demanding in terms of light-collecting areas, and ideally requires telescopes of several meters in diameter, distributed over several square kilometers. The technique also requires the measurement of photon arrival times with a precision better than one ns at each telescope over baselines extending to km distances. During the past five years, it has become possible to realize the technical requirements for SII using sparsely distributed arrays of Imaging Air Cherenkov Telescopes (IACTs), which have adequate optical properties, sufficiently large mirror areas, and telescope time available during the full Moon [2, 3, 4].

The modern implementation of the SII technique employs IACTs which operate essentially independently from each other while observing the same astronomical object [4]. During an observation, each telescope records the time variation of the light intensity from the object at sampling rates approaching 1 GHz, during a single, continuous exposure. The photon flux data-stream is stored in a large (10-20 TB) disk located at each telescope. Precise inter-telescope clock synchronization (<200 psec), required for the accurate computation of the two-photon cross-correlation function, is achieved using commercial fiber-optic cables connected to a central synchronization clock using a White Rabbit chipset [5]. Two-telescope and multi-telescope cross-correlations, necessary for computing the interferometric visibility for populating the Fourier image plane, are computed offline using an algorithm that can be implemented on a high-speed Field Programmable Gate Array (FPGA).

Because individual telescopes need only be connected by simple commercial fiber-optic cables, an observatory can be constructed with an arbitrary number of SII telescopes, and they can be spaced on the ground with any desired geometric (or random) pattern. In contrast, the physical requirements of ground-based VHE Cherenkov astronomy dictate that IACT arrays have an approximately periodic telescope spacing. Although beyond the scope of the development described in this APC white paper, we envision a future multi-kilometer scale array of 10-m class mobile IACT-type reflectors (the Very Large Intensity Interferometer, or VLII) whose locations can be rearranged to tailor the array's desired imaging resolution and field of view to specific astrophysical targets. VLII would complement the VLTI [6] observations by providing access to shorter wavelengths with higher angular resolution. VLII would operate in the visible waveband in a way analogous to reconfigurable multi-kilometer-scale radio/sub-mm arrays such as the VLA [7] or ALMA [8].

## Status of the Development and Demonstration of SII Technology

During the past five years, the development of SII technology for astronomical implementation has been pursued by several groups around the world using simulated thermal sources in the laboratory [9,10,11], and sky tests with 1-3 meter class telescopes [12,13,14,15,16], such as a twin 3-m test facility west of Salt Lake City (StarBase-Utah) [17]. Key technologies developed for multi-kilometer baseline SII interferometry include low-noise photodetectors and preamplifiers, high-speed data acquisition systems and multi-telescope correlators (both real-time and offline), and high resolution (< 1 nsec) clock synchronization between telescopes separated by distances approaching several kilometers. By Spring 2018, laboratory tests and tests at StarBase-Utah had demonstrated sufficient maturity of these technologies to enable SII instrumentation on arrays of 10-m class IACTs, such as the VERITAS VHE Gamma-ray Observatory (Amado, AZ).

Beginning in summer 2018, the VERITAS IACT observatory was augmented to demonstrate the ability of IACT arrays to perform SII observations [18]. This augmentation (called VERITAS-SII) included custom high-speed/low-noise focal plane instrumentation, high-speed (250 MHz), continuously sampling electronics, high-speed data (20 GBase-T) transport, and fast ( < 100 psec) inter-telescope time synchronization over distances greater than 100 m (Figure 2).  By spring 2019, the VERITAS-SII observatory had conclusively demonstrated its ability to measure intensity interferometric visibility over distance greater than 100m [19], allowing the measurement of the sub-milliarcsecond diameters of two bright stars in a  < 10 nm full-width waveband  centered about 415 nm [21,21]. These milestones have firmly established the feasibility of using IACT arrays with independent telescope data recording and offline software correlation for SII measurements.

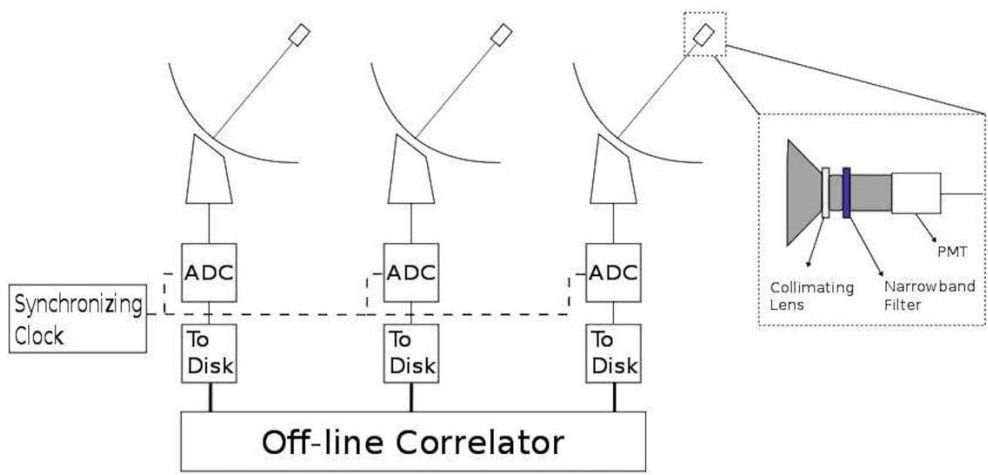

*Figure 2: Schematic of VERITAS-SII data recording system. Each telescope focal plane contains a narrowband optical filter, high-speed photodetector (photomultiplier tube - PMT), and low noise preamplifier which sends the continuous photoelectron currents to be digitized by a fast (250 Ms/s) ADC (FADC). The digitized currents are stored on high-speed RAID disks at each telescope, and the multi-telescope photon correlations are computed offline using an FPGA-based software correlator. The FADC clocks are synchronized using a commercially available White Rabbit-based fiber-optic clock distribution system[5].*

Additional improvements to VERITAS-SII are currently underway which will support a proposed SII observing campaign on 30 bright astrophysical targets during the 2019-2020 observing season, including hot (O/B/A) stars, rapid rotators, and binary systems.  The Spring 2019 VERITAS-SII observations have already exceeded the instantaneous visibility sensitivity of the pioneering NSII observatory.  It is expected that by summer 2020, VERITAS-SII will demonstrate the ability to fit astrophysical image models to the observed Fourier image plane, thereby demonstrating model-dependent SII imaging.  These observations will provide experience and data that can inform the design and simulation of the SII performance of a potential implementation of SII on the future Cherenkov Telescope Array (CTA).

CTA is a next-generation ground-based VHE gamma-ray observatory using the IACT array technique [22,23]. CTA is in an advanced design and prototyping phase, with construction and operation anticipated during the upcoming decade. The CTA Observatory (CTAO) design includes a Northern hemisphere site (CTA-N), to be located in the Canary Islands, and a Southern Hemisphere site (CTA-S), to be located in the Chilean Atacama Desert. CTA reaches its design goals by employing three basic sub-groups of telescopes: 4 Large-Sized Telescopes (LSTs), which utilize a 23 m diameter parabolic design at the center of the array for achieving the lowest gamma-ray energy threshold; an array of Medium-Sized Telescopes (MSTs) using a 12 m diameter modified Davis-Cotton design that is sensitive to the mid gamma-ray energy range, with 25 to be built in CTA-S and 15 in CTA-N; and in CTA-S, an array of 70 Small-Sized Telescopes (SSTs) that provides sensitivity to the highest energy gamma-rays, and which extends to baselines of a few km (Figure 3).

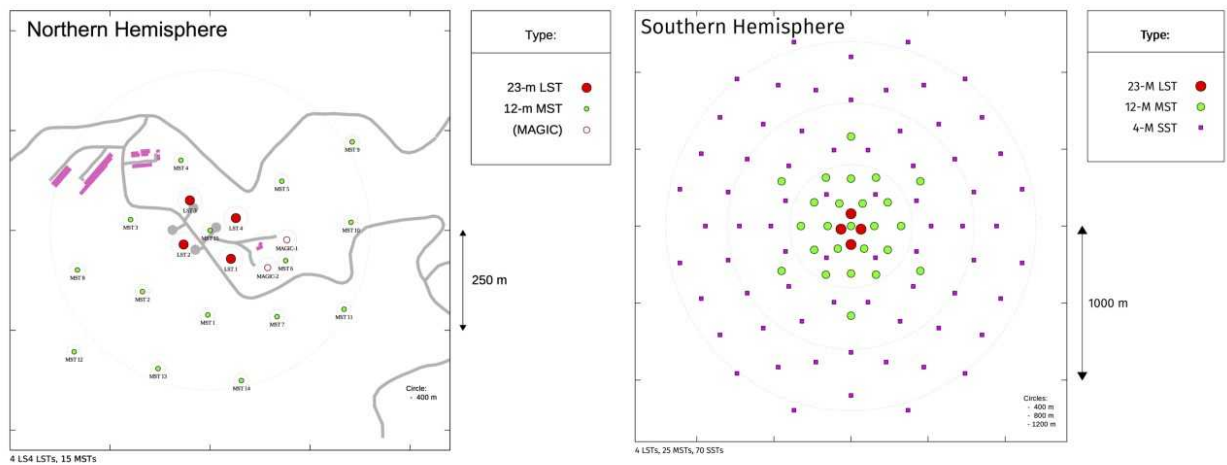

*Figure 3: Tentative telescope locations for the future CTA-N (left) and CTA-S (right) baseline array layout. The markers denote the ground locations of the various telescope types: red dots: LST; green circles: MST; purple squares; SST. Yellow circles indicate the location of the existing MAGIC telescopes.*

In the baseline configuration with nearly 100 telescopes spread over a few square km, the southern CTA array (CTA-S) will ultimately reach a combined light-collecting area of around 10,000 m$^2$. If CTA telescope baselines of 2 or 3 km are employed for intensity interferometry at short optical wavelengths, resolutions approaching 30 $\mu$as would be achievable for bright, hot stars. This visible band angular resolution is almost three orders of magnitude better than the Hubble Space Telescope, making CTA-SII the only astronomical facility capable of resolving visible wavelength features at the same resolution as long-baseline radio interferometry observatories such as achieved by intercontinental Very Long Baseline Interferometry (VLBI) [24] or the Event Horizon Telescope (EHT) [25] (Figure 4).

In Table 1, we compare the expected performance of an ideal Stellar Intensity Interferometer using several IACT observatories/telescopes with the relevant IACT mirror diameter, equipped with photodetectors with 35% quantum efficiency and 250 MHz/1 GHz bandwidth acquisition electronics. The sensitivity is calculated relative to the historic NSII observatory. It is clear that VERITAS-SII and either of the CTA arrays upgraded for SII has the potential to substantially exceed the performance of NSII.

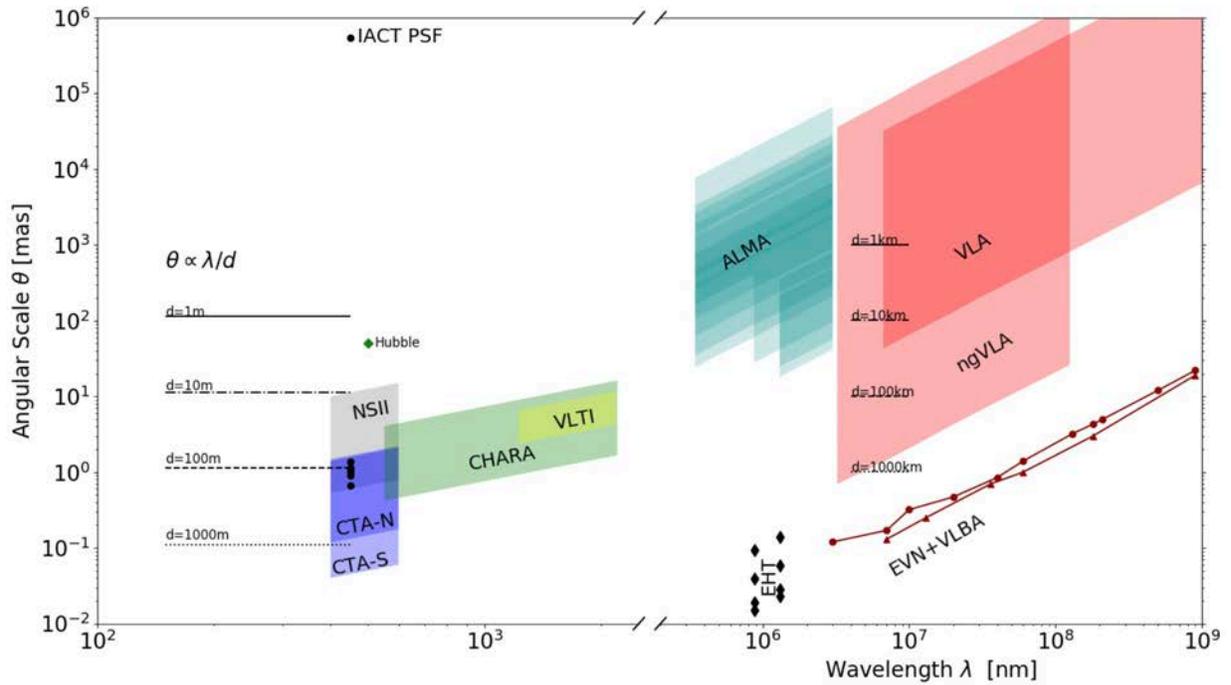

Figure 4: The angular resolution of various facilities in comparison to the resolutions that could be reached with the CTA telescopes equipped as intensity interferometers (CTA-SII). Round markers are for the VERITAS-SII implementation; diamond markers are for the Event Horizon Telescope interferometer [21].

| SII Instrument | D [m] | (S/N) cf NSII |
|---|---|---|
| NSII | 6 | 1.0 |
| StarBase-Utah | 3 | 0.84 |
| VERITAS-SII | 12 | 13.48 |
| LST | 23 | 98.94 |
| MST | 12 | 26.93 |
| SST | 4 | 2.99 |

Table 1: A comparison of current/future SII IACT telescope sensitivities with NSII. Starbase-Utah and VERITAS-SII both use 250 MHz detector bandwidth and 35% photon detection efficiency. The CTA detector bandwidth is assumed to be 1 GHz and its photon detection efficiency to be 35%.

Figure 5 shows the simulated best image resolution that could be achieved with CTA using the baseline locations of the SST and MST telescopes [26]. The simulation assumes an 8-hour continuous observation of a 0.8 mas diameter star about the star's culmination. The simulation assumes SII observation at 420 nm, and includes the telescope position sampling effects on the Fourier *u-v* image plane, but not the (currently unknown, system-dependent) electronics noise effects nor the algorithm-dependent phase reconstruction uncertainties. With a sufficiently bright star to ensure a high signal-to-noise ratio in the SII observations, a strikingly realistic image of the star and its surface features can be achieved. Once the final CTA telescope design choices are known, these simulations can be updated to include the system-dependent effects to accurately predict the final SII observatory resolution.

## Strategic Plan and Schedule

The path toward the realization of a modern imaging SII astronomy observatory will involve the development of instrument technology, the creation of a standard SII analysis pipeline, and the development of SII science planning toolkits.

Instrumentation technology
We have developed a framework of six observational milestones to provide a strategic ladder for developing the full scientific potential of imaging SII:

1. Demonstration of sensitivity to stars with $m_v \leq 2$.
2. Demonstration of sensitivity to stars with $m_v \leq 7$ (requires MST/LST).
3. Demonstration of angular resolution at ~1 mas scale.
4. Demonstration of angular resolution at ~100 μas scale.
5. Demonstration of model-dependent imaging.
6. Demonstration of model-independent imaging.

The achievement of the SII observational milestones is planned to occur in sequence in three distinct phases, with a schedule linked to the operations phase of VERITAS-SII and the construction phase of CTA. The three-stage approach will employ the use of commissioned facilities, such as VERITAS-SII, as well as SII on future instruments (i.e. CTA-SII). Some of the milestones will be achieved with partial construction of the final CTA array. All of these phases can be achieved in the science verification phase of CTA, and the first two phases can potentially be realized before the full CTA threshold array has been completed.

*Phase I - Proof of Principle*
This phase would be a 'state of the historic' system capable of matching the results of the NSII by achieving milestones 1 and 3. We remark that as of the date of this document (July 10, 2019), Phase I capability has already been demonstrated by the 2018-2019 VERITAS-SII instrumentation and observations. The VERITAS-SII Phase I results can be used to guide simulations of the performance of CTA-SII Phase I. A CTA-SII demonstration of Phase I would require an array of at least two functioning telescopes, preferably identical, separated by

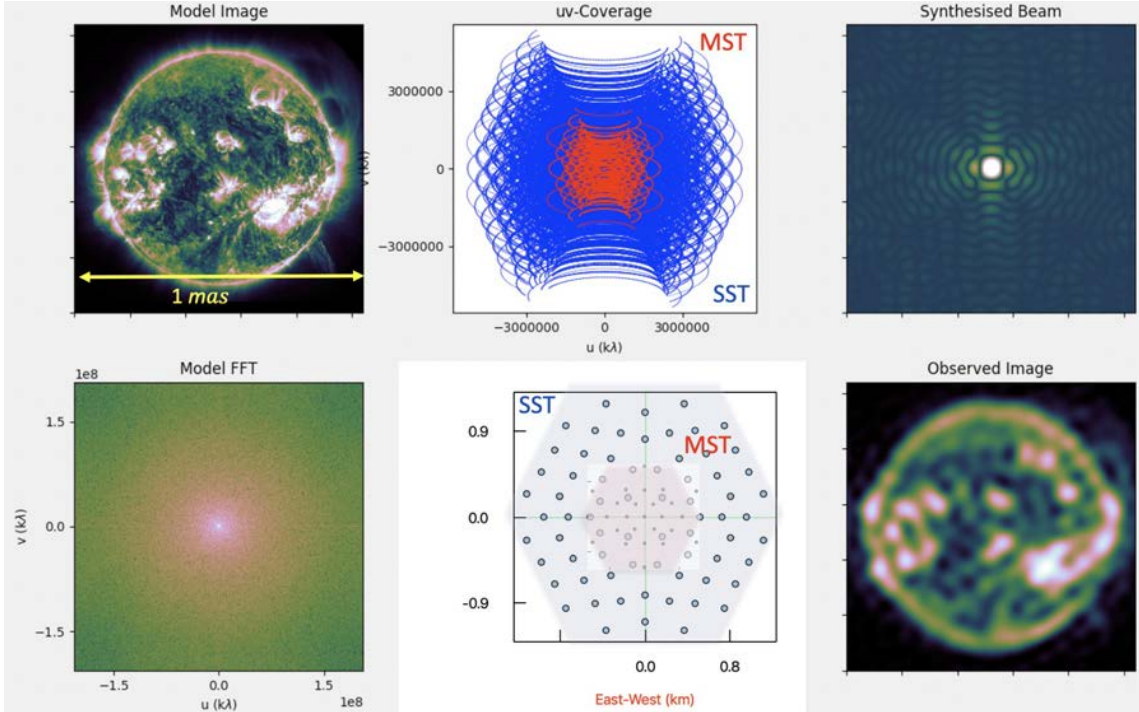

*Figure 5: Simulation of Fourier u-v image plane sampling effects on the synthetic SII image reconstruction for the combined baseline CTA SST/MST array [26]. **Upper left:** Input stellar image of a star; 0.8 mas diameter. **Upper middle:** Fourier u-v image plane coverage assuming 420 nm optical wavelength and ±4 hour exposure about the culmination of the star at 45 degrees above the horizon. **Bottom Middle:** Physical location of the simulated SSTs/MSTs in the CTA Array. **Bottom Right:** reconstructed image (N.B.: the simulation includes FFT sampling effects only; this does not include noise and phase reconstruction uncertainties).*

~100 m baselines. The accomplishment of CTA-SII Phase I may occur in 2022, depending upon the progress of CTAO construction.

*Phase II - State of the Current Technology*
The achievement of milestones 1, 2, 3, and 5 would herald an equivalent increase in sensitivity relative to the original NSII results on hot stars as realized by the current generation of amplitude interferometers (VLTI/CHARA/NPOI) [27,28,29] relative to the original Michelson interferometry on cool stars [30]. We remark that the VERITAS-SII observation program during 2019-2020 may potentially achieve Phase II by mid-2020. The VERITAS-SII Phase II results can be used to guide simulations of the performance of CTA-SII Phase II. A CTA demonstration of Phase II would necessarily occur later (early to mid-2020's), as it will require an array of at least five functioning MSTs.

*Phase III - State of the Art*
A state of the art SII facility, setting world-class standards for interferometric observations, will be realized by achieving all six SII technology milestones. Phase III cannot be achieved by VERITAS-SII alone due to the limited number of VERITAS telescopes. However, the results from VERITAS-SII and CTA-SII Phase II will guide the simulation of CTA- SII Phase III. CTA-SII Phase-III will only be achievable with the commissioning and operation of the threshold array of CTA (20 telescopes). The commissioning and operation of the CTA threshold array are expected to occur on or after the mid-2020s.

*Additional Technology Improvements*
The SII pathway outlined above assumes the simplest implementation of SII in CTA-SII: single photomultipliers with a single narrowband optical filter and I GHz sampling rate. Additional gains in signal-to-noise (S/N) may be achieved using higher quantum efficiency photosensors (e.g., SiPMs) [31,32] or instrumenting multiple, independent optical bandwidths [9] on a single telescope. In addition, SST/MST designs employing Schwartzschild-Couder optics possess better than 100 psec on-axis isochronicity [33], allowing an increase of the sampling rate up to 10 GHz. The sum of these technology improvements may potentially improve the telescope S/N ratio by a factor of 3-10 beyond the performance detailed in Table 1. These technology improvements will be pursued independently of, but coordinated with, the three-phase strategic approach outlined above.

Development of SII Analysis Pipeline
The massive nightly SII datasets require standardized cross-correlation algorithms that are highly optimized for rapid analysis. Issues such as calibration, excessive noise, and baseline drift must be automatically accounted for in the calculation of the interferometric visibility over each Fourier plane trajectory. The optimum analysis will certainly depend upon the S/N ratio, and so it will be necessary to develop different analyses for optimum sensitivity when combining different telescope types (e.g. SST-LST pairs). Similarly, successful image reconstruction will require implementation of various strategies for phase-recovery, model-dependent image reconstruction during Phase II, and the development of model-dependent imaging for Phase III. We envision a community SII pipeline analysis working group developing the analysis and image reconstruction pipeline independent of, but coordinated with, instrument technology development. A critical resource for the community SII pipeline analysis working group will be the establishment of an online public database of raw data from IACT array SII observations, with a common set of data access tools. This large (20-50 TB) database will be hosted at one or more high-performance computer facilities, preferably with GPU farms, and will be made available as an open sandbox for community development of SII algorithms and image reconstruction techniques.

Development of SII Science Planning Toolkits
The development of an SII science toolkit is an essential part of the strategic plan for the observatory. The SII science toolkit will allow arbitrary array configurations to be simulated, and will be linked to standard astronomical catalogs. The feasibility and timing of individual observations can be predicted by the observational toolkit. The toolkit will also calculate the number of hours required to achieve a specific resolution (e.g. relative error in stellar limb darkening coefficient) as a function of stellar temperature, magnitude, observatory latitude/longitude, observation wavelength, source sky position, and lunar illumination. This toolkit is essential for building a robust science observation plan at an IACT observatory. During Summer 2019, we have developed an SII science planning toolkit for use with VERITAS-SII during the 2019-2020 observing season [34]. We expect the toolkit will continue to be improved and updated based upon the instrumental response of the VERITAS-SII observatory. In future years, the toolkit will be updated by the SII community to incorporate the instrument response of CTA-SII.

## Organization, Partnerships, and Current Status

The development and implementation of SII technologies is proceeding through parallel efforts at multiple institutions in Europe, Asia, and the US. The VERITAS-SII development was funded by an NSF RAPID AST grant to the University of Utah for the purchase of hardware for the VERITAS-SII augmentation; VERITAS-SII observations are coordinated via the VERITAS Science Board and through formal observing proposals to the VERITAS Time Allocation Committee.

CTA-SII development is currently organized around individual CTA telescope/camera development projects; design and development of SII technology is in various stages of development for each of the CTA telescope designs (SST, MST, and LST). The CTA SII Science Working Group serves as a clearinghouse for CTA-SII technology development, as well as for coordination with VERITAS-SII and other non-CTA standalone technology development efforts. The organization, structure and current status of the CTA collaboration is described in detail in a separate APC white paper [35].

During 2018, the CTA SII Science Working Group commissioned a detailed internal study of a possible augmentation of CTA with SII. This comprehensive study explored the science case for ultra-high angular resolution optical astronomy at the sub-100 µas scale, reviewed the current state of the development of SII technology, and provided an estimation of the potential augmentation cost to create CTA-SII. The study, completed in May 2019, was distributed to the CTA Consortium Board for feedback and discussion. At the Lugano CTA Consortium (CTAC) meeting (June 2019), the CTAC Board approved a recommendation to CTAO for the inclusion of SII science and capability in CTA. The recommendation asked that CTAO assess design considerations for CTA infrastructure and other work packages that could allow a possible future augmentation of CTA for SII capability. A decision to formally include SII into CTA will be made after the completion of design and cost studies of individual telescopes, potential unification of individual telescope designs options, and availability of funding for SII augmentation from the relevant US and International funding agencies. The U.S. groups, represented by individuals in this APC White Paper, propose to make a substantial contribution to the completion of CTA-SII, in collaboration with international partners from CTAC.

VERITAS-SII and CTA-SII imaging observations in the visible wavelengths are strongly complementary to ongoing imaging observations at longer wavelengths (J, H, K) by amplitude interferometers, such as VLTI, CHARA, and NPOI. Consequently, the CTA-SII Working Group has begun discussions with these observatories to pursue strategies for synergistic multiwavelength imaging and observation, and potential joint research activities. As the technology for SII develops during the upcoming decade, we expect increasingly close collaboration and joint research projects between intensity and amplitude interferometric observatories.

## Cost Estimates

SII observations of bright stellar-like sources may be made under imperfect sky conditions, e.g. with drifting clouds, hazy skies, or even full Moon. This capability makes them complementary to the more demanding dark sky conditions necessary for VHE gamma-ray observations. The use of CTA-SII under moonlight conditions therefore represents a way of maximizing the science return from constructed telescope facilities when the conditions for VHE astronomy are not suitable. Because the CTA instrument will be an existing, fully funded instrument, the operation of CTA-SII will generate a 5-10% increase in the cost of observatory operations.

The capital costs of adding SII capabilities to CTAO is conservatively estimated to be $5-6M US, or approximately 1.5% of the CTA construction cost. This estimate includes the costs for externally mounted photodetectors and optics, data recording electronics, high speed data transport and synchronization, storage for nightly observations, and offline correlation and image reconstruction. Additional cost savings could be made if the SII instrumentation was internal to the different Cherenkov cameras. However, this possibility must be weighed against the requirement to control systematics by using consistent hardware between differing telescope types, and the freedom to upgrade the SII instrumentation with minimal impact on the IACT focal plane instrumentation.

A substantial driver of the estimated CTA-SII cost is the number of correlators needed to perform two-telescope correlations for every telescope pair in the array. By using offline, FPGA-based software correlation, the scale of this cost is substantially reduced, as the data from each telescope can be routed to an array of FPGA-based correlators through the high-speed network connections. The use of software correlators also provides the ability to compute higher-order correlations (e.g. three-telescope) between telescopes in the array which contain additional information that can be employed for phase recovery in image reconstruction [36,37].

We have examined the cost implications of the various potential Intensity Interferometry solutions for CTA-SII. Current numbers represent best estimates based upon the use of readily available commercial off-the-shelf equipment. Significant cost savings (a factor of 2-5) could be realized by using custom built hardware components. This effort would have significant development costs that would be economical only when amortized over a large number of telescopes (>10). The lessons learned from the laboratory and the on-sky observations during Phase I and Phase II will be critical for optimizing the hardware design and solidifying the concomitant cost estimates for the final implementation of CTA-SII.

## Summary

With the implementation of multi-kilometer baseline Stellar Intensity Interferometry, the next decade will usher in an era of visible-band interferometry and imaging in the unexplored 30-40 µas angular resolution regime, leading to breakthrough science opportunities [38]. Future advances may lead to the construction of a major visible waveband imaging observatory array, with a similar legacy science impact of other major interferometric astronomical imaging observatories.